\pdfoutput=1
\documentclass[runningheads]{llncs}

\usepackage[T1]{fontenc}
\usepackage{graphicx}
\usepackage{booktabs}
\usepackage{amsmath}
\usepackage{xcolor}
\usepackage{hyperref}
\usepackage{float}
\usepackage{etoolbox}
\AtBeginEnvironment{thebibliography}{\footnotesize\let\small\relax\setlength{\itemsep}{0pt}\setlength{\parskip}{0pt}}

\makeatletter
\renewcommand\section{\@startsection{section}{1}{\z@}%
  {-12pt plus -2pt minus -1pt}{6pt plus 1pt}{\normalfont\large\bfseries\boldmath\raggedright}}
\renewcommand\subsection{\@startsection{subsection}{2}{\z@}%
  {-10pt plus -2pt minus -1pt}{4pt plus 1pt}{\normalfont\normalsize\bfseries\boldmath\raggedright}}
\renewcommand\paragraph{\@startsection{paragraph}{4}{\z@}%
  {4pt plus 1pt}{-0.5em}{\normalfont\normalsize\itshape}}
\makeatother

\graphicspath{{figs/}}

\begin{document}

\title{"If It Looks Like a User": Measuring Real-Time Moderation Effects via Social Media Simulation}
\titlerunning{Real-Time Moderation Effects via Social Media Simulation}

\author{Enrico~Verdolotti\inst{1,2}\orcidID{0009-0009-0330-3537} \and\\ Gianluca~Nogara\inst{2}\orcidID{0000-0002-4412-131X} \and\\
  Luca~Luceri\inst{3}\orcidID{0000-0001-5267-7484} \and\\ Silvia~Giordano\inst{2}\orcidID{0000-0003-2603-9029}}
\authorrunning{E. Verdolotti et al.}
\institute{Universit\`a della Svizzera italiana (USI), Lugano, Switzerland\\
  \email{enrico.verdolotti@usi.ch}
  \and
  SUPSI, DTI-ISIN, Viganello, Switzerland\\
  \email{\{enrico.verdolotti, gianluca.nogara, silvia.giordano\}@supsi.ch}
  \and
  USC Information Sciences Institute, Marina Del Rey, CA, USA\\
  \email{lluceri@isi.edu}}

\maketitle
\vspace{-5mm}

\begin{abstract}
Agent-based social media simulators offer a controlled environment to study content moderation, yet their value hinges on how faithfully they reproduce real platform dynamics.
We develop a calibrated extension of SimSoM, an agent-based model of information diffusion on social networks, grounded in a real-world dataset of online vaccine discourse during the COVID-19 pandemic.
Our approach replaces ad-hoc parametrisations with empirically fitted distributions, optimised via CMA-ES (Covariance Matrix Adaptation Evolution Strategy) and validated against real data across temporal, distributional, and structural dimensions.
Using this validated simulator, we provide three key contributions.
First, we show that the calibrated model reproduces key statistical signatures of the empirical data, including activity distributions, post/reshare ratios, and temporal patterns.
Second, we apply established misinformation-spreader detection and prevention methods to both empirical and simulated data, progressively removing top-ranked users and showing that the resulting decline in low-quality content is consistent across the two.
Third, comparing static (retroactive) and dynamic (in-simulation) moderation across 30 network realisations, we show that static evaluation significantly overestimates the effectiveness of user bans for the most effective detectors: when moderation is applied in real time, compensatory resharing by the remaining users dampens the expected reduction in low-quality content, so static estimates should be read as an upper bound.
These findings highlight the necessity of simulation-based evaluation for content moderation policies and contribute a reusable, empirically grounded simulation framework.

\end{abstract}

\section{Introduction}
\label{sec:intro}

Online misinformation poses a persistent threat to public discourse, particularly during crises such as the COVID-19 pandemic~\cite{lazer2018science,vosoughi2018spread}, yet evaluating the effectiveness of interventions remains fundamentally unresolved~\cite{truong2025delayed,cima2024greatban}.
Social media platforms have responded with a range of moderation interventions, from content removal and demotion to account suspension and deplatforming~\cite{jahn2025friction,schneider2023moderation,chandrasekharan2017youcantstay}.
Most empirical assessments of moderation rely on \emph{static, retroactive} analysis: given a historical dataset, one identifies the most harmful accounts, removes their activity, and measures the resulting improvement in information quality~\cite{deverna2024superspreaders,verdolotti2025spreaders,nogara2024misinfo}.
While informative, this approach ignores the feedback loops that arise when moderation occurs \emph{in real time}: the remaining network adapts, compensatory behaviours emerge, and the information ecosystem reshapes itself around the intervention~\cite{hortaribeiro2021migrations,cima2024greatban,cinelli2021echo}.

In contrast, agent-based simulation can capture adaptive responses to moderation: by embedding policies within a running simulation, one observes emergent reactions that static analysis, by construction, cannot. A simulator is only as credible as its calibration, however; as Box noted, ``all models are wrong, but some are useful''~\cite{box1976science}, and a social media simulator's usefulness depends on how well its dynamics mirror empirical reality.

In this paper we present a calibrated, empirically validated extension of SimSoM~\cite{truong2024simsom}, an agent-based model of information diffusion and manipulation on social networks.
We ground the simulator in a publicly available longitudinal dataset of Italian-language COVID-19 vaccine discourse~\cite{pierri2021vaccinitaly,nogara2024misinfo}, and replace the original model's stylised parametrisation with distributions fitted to observed data.
In particular, user activity levels are drawn from a Generalized Beta of the Second Kind (GB2) distribution~\cite{mcdonald1995gb2} whose parameters are optimised via CMA-ES~\cite{hansen2016cmaes}, and content quality follows a bimodal Beta mixture calibrated by grid search on the empirical credibility distribution.

We address three research questions.
\begin{itemize}
    \item \textbf{RQ1}: \emph{Can the calibrated SimSoM reproduce the statistical signatures of real platform dynamics?}
    \item \textbf{RQ2}: \emph{Does the evaluation of moderation strategies on simulated data yield effects consistent with those measured on empirical data?}
    \item \textbf{RQ3}: \emph{How does the effect of a moderation intervention applied dynamically (in-simulation, allowing the network to react) compare with its static (retroactive) evaluation?}
\end{itemize}
The intervention itself is a single, one-shot user ban (Section~\ref{sec:rq3}); what is \emph{dynamic} in RQ3 is not the policy but the agents' adaptive response to it, which static removal cannot capture. We focus on \emph{user-centred} moderation (account bans), rather than content-centred actions (labelling, demotion): account-level interventions such as suspension and deplatforming are among the most consequential, widely deployed levers, and are precisely where the compensatory network responses we study arise.

Our contributions are threefold. First, we propose an empirically grounded calibration and validation methodology for social media simulators: user activity is fitted to a
GB2 distribution via CMA-ES, content quality follows a bimodal Beta mixture calibrated by grid search, and validation spans temporal, distributional, and structural dimensions. We show that the calibrated model closely reproduces activity distributions, post/reshare ratios, and quality profiles of the empirical data~(\textbf{RQ1}).
Second, we provide evidence that simulated data produced by the
calibrated model can reliably serve as a proxy for evaluating moderation strategies: four independent detection methods~\cite{verdolotti2025spreaders} yield comparable user-removal effectiveness curves on empirical and simulated data, with confidence intervals consistently overlapping~(\textbf{RQ2}).
Third, using a paired comparison over 30 network realisations, we show that static, retroactive evaluation significantly overestimates the impact of user bans for the effective detectors (and not for the ineffective ones): when moderation is applied dynamically, compensatory resharing by the remaining users dampens the expected reduction in low-quality content, so static evaluation acts as an upper bound~(\textbf{RQ3}).

\paragraph{Relation to prior work.}
From our recent work on misinformation-spreader ranking~\cite{verdolotti2025spreaders} we inherit only the \emph{inputs} to RQ2: the four spreader-ranking methods and the VaccinItaly credibility-annotation pipeline, reused unchanged as off-the-shelf detectors. Everything else is novel: the calibrated SimSoM extension (GB2 activity via CMA-ES, the bimodal-Beta quality mixture, the correlation/appeal structure), the validation protocol of Section~\ref{sec:rq1}, and, most importantly, the static-vs-dynamic comparison of Section~\ref{sec:rq3}. In short, \cite{verdolotti2025spreaders} provides \emph{whom} to remove; this paper measures \emph{what happens} when they are removed in real time.

\section{Related Work}
\label{sec:related}

\paragraph{Social media simulation.}
Agent-based models have a long tradition in computational social science~\cite{gilbert2005simulation,epstein2006generative}.
SimSoM~\cite{truong2024simsom} introduced a minimalistic model of information diffusion on directed networks, demonstrating how adversarial tactics (infiltration, flooding, deception, and coordinated inauthentic behaviour) degrade information quality.
More recent work has extended agent-based modelling to cross-platform diffusion~\cite{murdock2024crossplatform}, language-sensitive cloning of online communities~\cite{puri2024digital}, and LLM-powered content generation~\cite{liu2025mosaic}.
However, existing simulators typically rely on stylised parameter choices, are rarely validated against real-world intervention outcomes, and almost exclusively evaluate moderation in static, retrospective settings. 
In contrast, we develop a simulator that is fully calibrated on empirical data, validate it against empirical targeted-removal experiments, and use it to directly compare static and dynamic moderation regimes.

\paragraph{Misinformation spreading and detection.}
The empirical study of misinformation has established that false content spreads faster and more broadly than truthful information~\cite{vosoughi2018spread}, and that echo chambers amplify polarisation~\cite{cinelli2021echo}.
Identifying the most influential misinformation spreaders is a prerequisite for effective moderation.
Verdolotti et al.~\cite{verdolotti2025spreaders} proposed a behavioural-archetype framework distinguishing super-spreaders, amplifiers, and coordinated accounts, introducing time-aware ranking methods such as the TASH-Index that outperform static centrality measures.
Coordinated inauthentic behaviour detection~\cite{pacheco2021cib,cresci2020decade,luceri2024unmasking,luceri2021bots} complements individual-level ranking by identifying organised campaigns and automated accounts.

\paragraph{Content moderation.}
Schneider and Rizoiu~\cite{schneider2023moderation} provided empirical evidence that moderation is effective at reducing harmful content exposure, but noted substantial variation across platforms and policy designs.
Jahn et al.~\cite{jahn2025friction} studied friction-based interventions within a simulation framework, showing that small design nudges can curb misinformation spread.
Truong et al.~\cite{truong2025delayed} audited takedown delays across major platforms and demonstrated that even moderate delays drastically reduce moderation effectiveness.
Network dismantling~\cite{ren2019dismantling}, i.e.\ the progressive removal of nodes to maximally disrupt a network, provides the mathematical backbone for evaluating targeted user removal strategies.

\paragraph{Simulation validation.}
Validating agent-based models against empirical data is a well-recognised methodological challenge~\cite{windrum2007empirical,collins2024validation}.
Common approaches include distributional comparison, temporal pattern matching, and sensitivity analysis.
Our work extends this tradition by proposing a multi-dimensional validation protocol that jointly assesses activity distributions, temporal dynamics, content composition, and moderation outcomes.

\section{Data and Simulation Framework}
\label{sec:data_sim}

\subsection{Empirical Data: VaccinItaly}
\label{sec:data}

We ground our simulator in VaccinItaly~\cite{pierri2021vaccinitaly,nogara2024misinfo}, a longitudinal dataset of Italian-language COVID-19 vaccine discourse collected from December 2020 to November 2021 ($\approx306$ days).
VaccinItaly tracks public conversations about vaccines on Twitter, 
gathered through the streaming API with a curated, routinely updated list of Italian vaccine-related keywords and hashtags (e.g.\ \emph{vaccino}, \emph{AstraZeneca}, \emph{Pfizer}, \emph{greenpass}) so as to track trending terminology over the campaign; we use the Twitter posts and retweets, which map naturally onto the simulator's \emph{post} and \emph{reshare} actions.
The resulting subset comprises 819{,}952 actions (posts and reshares) by 74{,}234 unique users, annotated with per-item credibility scores derived from source reliability ratings.
Content credibility is determined by extracting URLs from each post and querying the NewsGuard reliability index of the associated domains, following the methodology of DeVerna et al.~\cite{deverna2024superspreaders} and Verdolotti et al.~\cite{verdolotti2025spreaders}.
We adopt a normalised credibility threshold of 0.39 to distinguish low-quality (LQ) from high-quality content.\footnote{Throughout this paper, we use \emph{low-quality} as a generalisation of \emph{misinformation}: while prior work on VaccinItaly focused specifically on vaccine misinformation~\cite{verdolotti2025spreaders,deverna2024superspreaders}, our simulator and evaluation framework apply to any content quality dimension.}

\subsection{SimSoM and Our Extensions}
\label{sec:simsom}

SimSoM~\cite{truong2024simsom} is an agent-based model in which users interact on a directed follower network.
At each time step, active users either create original posts or reshare content from their feed.
A recommender system populates each user's feed with a mixture of in-network (friends' posts) and out-of-network (trending) content, ranked by appeal and recency.
The original model uses a power-law activity distribution and uniform quality assignment.

We extend SimSoM along four axes, replacing stylised assumptions with empirically calibrated mechanisms:

\paragraph{(i) Activity generation via GB2.}
We model per-user mean activity with a Generalized Beta of the Second Kind (GB2) distribution~\cite{mcdonald1995gb2}, constructed as $X = b \cdot (U/(1-U))^{1/a}$ where $U \sim \text{Beta}(p, q)$.
The four parameters $(a, b, p, q)$ control tail heaviness, scale, mass near zero (inactive users), and tail decay, respectively.
We optimise these jointly with four temporal parameters (user inertia, user variability, global inertia, global variability) using CMA-ES~\cite{hansen2016cmaes} via Optuna~\cite{akiba2019optuna}, minimising a composite fitness function over percentile errors, autocorrelation, and daily totals on VaccinItaly.
CMA-ES (Covariance Matrix Adaptation Evolution Strategy) is a derivative-free, population-based optimiser that samples candidate parameter vectors from a multivariate Gaussian and iteratively adapts its mean and covariance towards the better candidates, making it well suited to our non-convex, noisy, gradient-free fitness landscape.
The fitted values are $a{=}0.472$, $b{=}0.008$, $p{=}2.439$, $q{=}3.106$.

Daily activity counts are then sampled as $n(t) \sim \text{Poisson}(\bar{\lambda} \cdot s_u(t) \cdot s_g(t))$, where $\bar{\lambda}$ is the GB2-drawn mean activity rate, and $s_u(t)$, $s_g(t)$ are user-level and global temporal state multipliers evolving as random walks with calibrated inertia.

\paragraph{(ii) Quality distribution.}
Content quality is drawn from a bimodal Beta mixture: 18\% of users sample from a low-quality component ($\mu{=}0.14$, spread~12) and 82\% from a high-quality component ($\mu{=}0.84$, spread~8).
Parameters were selected by grid search over 5{,}040 combinations, minimising weighted error on per-action quality fractions from VaccinItaly.

\paragraph{(iii) Correlation structure.}
We induce empirically motivated correlations via the Iman-Conover method~\cite{iman1982distribution}: activity$\leftrightarrow$in-degree ($\rho{=}0.6$), activity$\leftrightarrow$quality ($\rho{=}{-}0.14$), and post-propensity$\leftrightarrow$activity ($\rho{=}0.13$).
Action type probabilities (post vs.\ reshare) follow a Dirichlet distribution with $\alpha{=}(0.075, 0.095)$, yielding the bimodal specialisation pattern observed in real data.

\paragraph{(iv) Content appeal and critical filtering.}
User-level content appeal is rank-correlated with in-degree via a right-skewed transform.
Low-quality content receives a sensationalism variance boost (factor 2.5 for $q < 0.35$), in line with the empirical observation that a small fraction of misinformation achieves disproportionate virality.
A critical filter governs reshare selectivity: extreme users (very low or very high quality) preferentially reshare quality-similar content, while moderate users are less selective ($\text{selectivity}{=}0.97$, $\text{extremism\_weight}{=}0.25$).

\paragraph{Network and experimental configurations.}
All simulations run on directed networks with average out-degree~3, loaded from pre-generated graphs consistent with empirical degree distributions, and span 306~days to match the VaccinItaly observation window.
We employ two network configurations for different experimental purposes.
For statistical validation and dismantling comparison (RQ1--RQ2), we use 90{,}000-node networks.
The size is dictated by the GB2 mass near zero: a non-negligible fraction of agents draws an activity rate so low that they perform no action during the 306-day window (\emph{inactive lurkers}).
Since VaccinItaly by construction only contains users who posted at least once, calibrating on a 74k-node graph would yield a systematically smaller \emph{active} population.
We therefore pick the smallest network size whose expected active population matches the empirical 74k under the calibrated GB2: 90k nodes (cf.\ Table~\ref{tab:comparison}, ``Total users'': $72.9\text{k}\pm 1.6\text{k}$).
Lurkers do not affect dismantling: by construction, they cannot appear in any LQ reshare network.
Each of the 50 runs generates a fresh network, so results are robust to network realisation.
For the static-vs-dynamic moderation comparison (RQ3), we use 30 independently generated 50{,}000-node networks with the same calibrated parameters, as a controlled substrate whose
fidelity is already established at 90k (Section~\ref{sec:rq1}). Each network is run through every condition (no moderation and each ranking method), so the static and dynamic regimes of a given method are evaluated on the \emph{same} topology and compared \emph{pairwise}, with the network realisation as the matching unit. This design removes the (large) between-network variance while still probing robustness across 30 realisations.

\section{Simulating Real Data (RQ1)}
\label{sec:rq1}

We validate the calibrated simulator by comparing 50 independent simulation runs, each on a freshly generated 90k-node network, against VaccinItaly across multiple statistical dimensions, to assess whether it reproduces the empirical signatures relevant to information diffusion and moderation.
Table~\ref{tab:comparison} summarises the key metrics.

\begin{table}[!htb]
\centering
\caption{Summary statistics for real and simulated datasets. \emph{Dataset statistics}: total actions (posts and reshares), unique users, observation window. \emph{Content composition}: \emph{orphan rate}~=~fraction of reshares whose parent post is absent from the collected data; \emph{post}/\emph{reshare fraction}~=~share of actions that are posts vs.\ reshares; \emph{unique targets ratio}~=~fraction of distinct content items ever reshared. \emph{User activity}: per-user action counts (median, p90), activity inequality (Gini), burstiness~\cite{goh2008burstiness} of inter-event times, and median \emph{response time}~=~delay between a post and its first reshare (see the sequential-scheduling discussion in the text). Simulation values are mean\,$\pm$\,std across 50 independent 90k-node runs.}
\label{tab:comparison}
\begin{tabular*}{\textwidth}{@{\extracolsep{\fill}}lr r@{}}
\toprule
\textbf{Metric} & \textbf{Real} & \textbf{Simulated} \\
\midrule
\multicolumn{3}{@{}l}{\textit{Dataset statistics}} \\
\quad Total actions & 819,952 & 973,539 $\pm$ 73,558 \\
\quad Total users & 74,234 & 72,915 $\pm$ 1,605 \\
\quad Duration (days) & 306.5 & 306.0 $\pm$ 0.0 \\
\addlinespace[2pt]
\multicolumn{3}{@{}l}{\textit{Content composition}} \\
\quad Orphan rate & 0.114 & 0.000 $\pm$ 0.000 \\
\quad Post fraction & 0.547 & 0.577 $\pm$ 0.011 \\
\quad Reshare fraction & 0.453 & 0.423 $\pm$ 0.011 \\
\quad Unique targets ratio & 0.122 & 0.113 $\pm$ 0.005 \\
\addlinespace[2pt]
\multicolumn{3}{@{}l}{\textit{User activity}} \\
\quad Actions/user (median) & 1.000 & 2.980 $\pm$ 0.141 \\
\quad Actions/user (p90) & 14.000 & 21.480 $\pm$ 1.199 \\
\quad Gini coefficient & 0.834 & 0.788 $\pm$ 0.006 \\
\quad Burstiness & 0.521 & 0.419 $\pm$ 0.009 \\
\quad Response time, h (median) & 0.106 & 4.872 $\pm$ 0.177 \\
\bottomrule
\end{tabular*}
\end{table}

\begin{figure}[!htb]
\centering
\includegraphics[width=\textwidth]{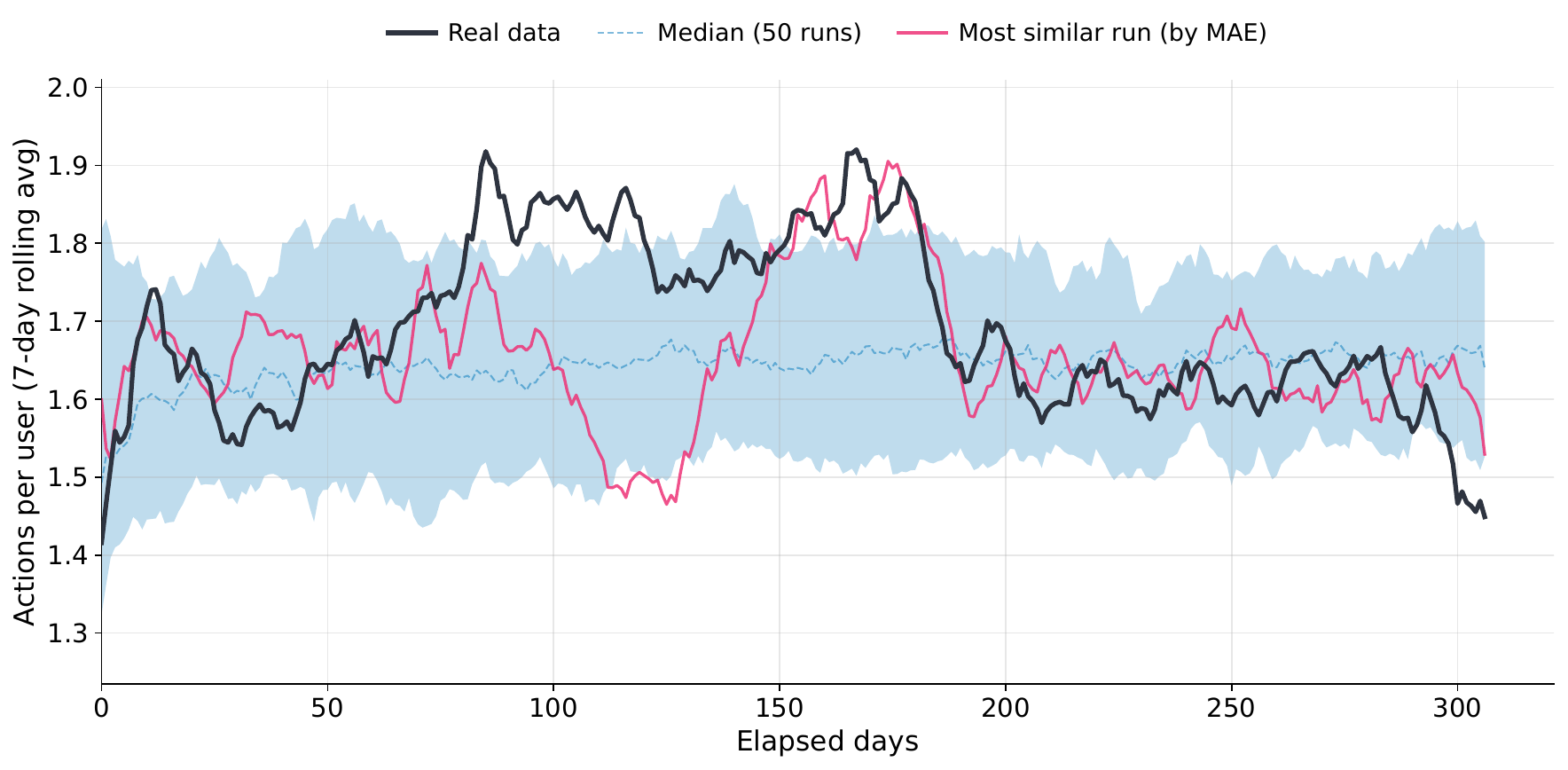}
\caption{Temporal comparison: 7-day rolling average of actions per user. Real series (dark), median across the 50 runs (dashed) with its 5th--95th percentile envelope (shaded band), and the single run closest to the real series by mean absolute error (solid).}
\label{fig:temporal}
\end{figure}

\paragraph{Temporal dynamics.}
Figure~\ref{fig:temporal} shows the 7-day rolling average of actions per user for VaccinItaly and the simulated runs (per-user normalisation makes the 90k-node simulation and the 74k-user dataset directly comparable).
To convey both typical and best-case behaviour, the figure overlays three curves: the real series; the \emph{median} across the 50 runs together with its 5th--95th percentile envelope (a robust, outlier-resistant summary of the run-to-run spread, not a confidence interval on the mean, and hence wide by design); and the single simulation run closest to the real series, selected as the one with smallest mean absolute error (MAE).
The aim is qualitative: rather than matching the exact trajectory, we ask whether the simulator \emph{endogenously} produces plausible dynamics. Two facts support this: the empirical series stays within the run-to-run envelope for essentially the whole window (so it is a plausible draw, not an outlier), and the closest run tracks it over long stretches, reproducing \emph{sustained} bursts of comparable magnitude (e.g.\ days~150--185) without being a lucky aberration, as it lies inside the band. The simulated temporal autocorrelation matches the real data within 2\% at lags 1--5.
The largest residual departures, most visibly around days~80--130, correspond to \emph{exogenous}, event-driven surges in the real discourse, e.g.\ the temporary suspension of the AstraZeneca vaccine in March~2021, the spring vaccination-campaign ramp-up, and the introduction of the ``Green Pass'' in summer~2021. These spikes are driven by external news shocks that lie outside the model's generative mechanism (the simulator has no exogenous-event channel), so neither the median nor any single run is expected to time-align with them; the calibrated dynamics track the baseline trend and its dispersion but, by construction, do not reproduce individual real-world events. This gap concerns the \emph{timing} of bursts, not the daily-aggregated quantities (activity level, post/reshare composition, LQ fraction) that drive the moderation analysis in Section~\ref{sec:rq3}.

\paragraph{Activity distributions.}
The simulated per-user activity distribution preserves the heavy-tailed character of the empirical data (Gini 0.788 vs.\ 0.834).
The discrepancy at the median (2.98 vs.\ 1.0 actions/user) reflects the simulator's fully-connected observation window: all generated actions are recorded, whereas the real dataset only captures actions matching the collection query.
Post and reshare fractions are closely matched (simulated: 57.7\%/42.3\% vs.\ real: 54.7\%/45.3\%).
The one notable gap is the median response time (4.9\,h simulated vs.\ 0.1\,h real, i.e.\ the delay between a post and its first reshare): the simulator's sequential scheduling prevents instantaneous within-tick reactions, inflating short intervals.
Since moderation is evaluated through daily LQ-fraction changes over a 120-day window, this affects sub-daily timing, not the aggregate quantities used in the moderation experiments of Section~\ref{sec:rq3}.

\paragraph{Quality profile.}
The per-action credibility distribution is bimodal in both datasets, with aligned low- and high-quality peaks and matching mass (mean and std differ by ${<}1$\,pt on the [0--100] scale): real and simulated data agree on the quantities that drive moderation analysis (bimodal shape, LQ/HQ mass ratio, and the LQ-fraction time series of Section~\ref{sec:rq3}). The real distribution is discrete (NewsGuard assigns one credibility score per domain), whereas the simulator samples a continuous Beta mixture.

\paragraph{Content composition.}
The unique targets ratio is closely matched (0.113 vs.\ 0.122); the simulated orphan rate is zero by construction (the simulator tracks all content internally), while VaccinItaly shows 11.4\% orphan reshares due to incomplete data collection. Overall, the calibrated simulator reproduces the essential statistical properties that directly govern information diffusion and moderation outcomes of VaccinItaly, supporting its use as a testbed for moderation experiments.

\section{Moderation Strategies: Real vs.\ Simulated Data (RQ2)}
\label{sec:rq2}

Having established statistical fidelity, we ask whether moderation strategies evaluated on simulated data yield effects consistent with empirical data. We apply four misinformation-spreader ranking methods (TASH-Index, Random Forest, Coordination Centrality, Repost Count~\cite{verdolotti2025spreaders}) to both VaccinItaly and the 50 simulated runs, then perform targeted user removal (network dismantling)~\cite{deverna2024superspreaders}, testing whether the simulator concentrates influence on the same users as real data.

\paragraph{Ranking methods.}
TASH-Index~\cite{verdolotti2025spreaders} is a time-aware social h-index weighting recent activity via exponential moving averages; Random Forest combines archetype features~\cite{verdolotti2025spreaders} in a supervised model; Coordination Centrality uses eigenvector centrality on a TF-IDF resharing-similarity graph~\cite{pacheco2021cib}; Repost Count is a frequency baseline on low-credibility reshares.

\paragraph{Dismantling protocol.}
Following~\cite{deverna2024superspreaders,verdolotti2025spreaders}, we build the LQ reshare network (an edge $u\!\to\!v$ of weight~$w$ means $v$ reshared low-quality content authored by~$u$ $w$ times), remove users in ranked order, and track the remaining LQ reshare weight~\cite{ren2019dismantling}; simulated curves are median $\pm$ 5th/95th percentile across the 50 runs.

\begin{figure}[!htb]
\centering
\includegraphics[width=0.99\textwidth]{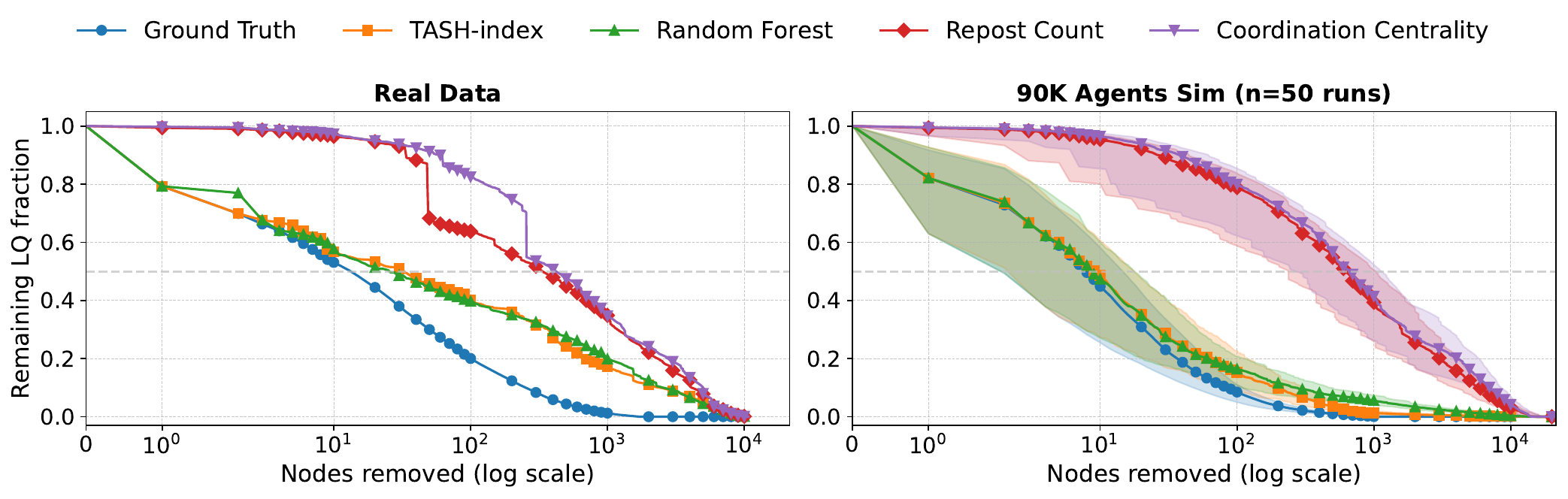}
\caption{Targeted user removal: remaining fraction of LQ reshare weight as users are removed one by one in ranked order, for real (solid) and simulated (shaded CI) data.}
\label{fig:dismantling_curves}
\end{figure}

\paragraph{Removal curves.}
Figure~\ref{fig:dismantling_curves} shows the dismantling of the LQ reshare network. Across all four methods, the shape and rate of decline match closely between VaccinItaly and simulation, with simulated CIs overlapping the real-data curve, consistent with the simulator reproducing the structural concentration of misinformation spreading and thus the relative effectiveness of targeted user removal.

\begin{figure}[!htb]
\centering
\includegraphics[width=0.99\textwidth]{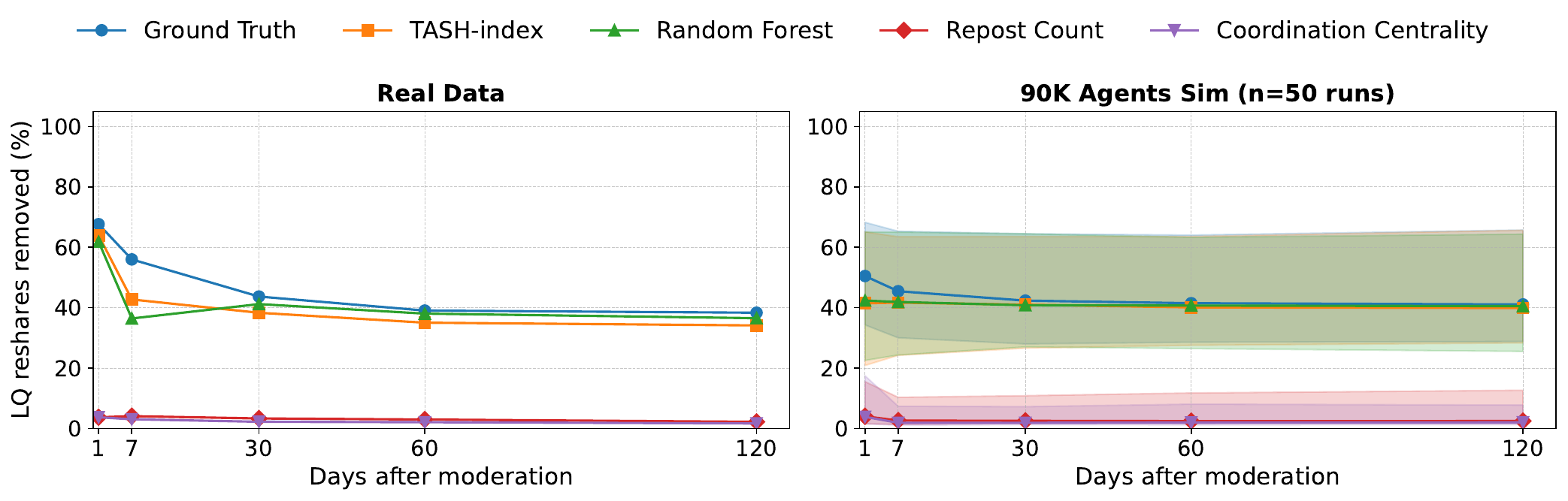}
\caption{Temporal effectiveness of banning the top-5 ranked users: percentage of LQ reshares attributable to banned users at increasing days after moderation, for real (left) and simulated (right, with CI) data.}
\label{fig:effectiveness}
\end{figure}

\paragraph{Temporal effectiveness.}
A complementary question is how the top-$k$ users historically contributed to LQ reshares at different time points in the dataset.
Figure~\ref{fig:effectiveness} fixes the banned set (top~5 by each ranking) and measures the percentage of LQ reshares attributable to them at 1, 7, 30, 60, and 120~days after a reference moderation day
(day 186, the same split used in RQ3).
On real data, the ground-truth drops from ${\sim}$65\% at day~1 to ${\sim}$40\% at day~120, as other users gradually compensate.
TASH-Index and Random Forest track this decay closely; Repost Count and Coordination Centrality remain near zero, confirming their inability to capture the dominant spreading structure.
The simulated data reproduces this temporal decay, with confidence intervals overlapping the real-data curves at every horizon.

These results validate the simulated data as a reliable proxy for evaluating moderation strategies: the simulator preserves not only aggregate statistics but also the structural and temporal consistency that determine which users are most influential in LQ content networks.

\section{Static vs.\ Dynamic Moderation (RQ3)}
\label{sec:rq3}

The central question is whether \emph{static} moderation evaluation (removing users retroactively from a completed dataset) faithfully predicts the effect of \emph{dynamic} moderation, where the remaining users adapt their resharing in response to the intervention, which static post hoc removal cannot capture.
We compare the two regimes across 30 independent 50k-node networks (Section~\ref{sec:simsom}). Because both regimes of a method run on the \emph{same} network, we compare them \emph{pairwise} (network as the matching unit), which cancels the large between-network variance and isolates the moderation effect.

\paragraph{Experimental setup.}
We run each simulation for 306 days with dynamic moderation activated at day~186 (approximately the 60/40 pre/post split in VaccinItaly).
At the moderation point, the top-5 users identified by each ranking method are permanently banned: their activity rate is set to zero, and their content is excluded from all feeds.
For the static condition, we take each network's unmoderated run and retroactively remove the top-5 users' post-moderation actions.
For each method, we report the change in mean daily LQ fraction ($\Delta$) under each regime and the within-network gap (static$-$dynamic) with a 95\% bootstrap confidence interval~\cite{efron1994introduction} and a Wilcoxon signed-rank test~\cite{wilcoxon1945individual} over the 30 networks.

\paragraph{Content-level comparison.}
Table~\ref{tab:static_vs_dynamic} reports the paired static$-$dynamic gap in mean daily LQ fraction across the 30 networks.
\begin{table}[!ht]
  \centering
  \caption{Static (S) vs.\ dynamic (D) moderation across 30 independent 50k-node networks (moderation at day~186, top-5 banned). $\Delta$: change in mean daily LQ fraction ($\downarrow$~=~less misinformation). Gap~=~S$-$D with a 95\% bootstrap CI over networks; a \emph{negative} gap means static overestimates real-time effectiveness (an upper bound). $p$: Wilcoxon signed-rank test (H$_0$: median gap~=~0). \textbf{Bold}: significant at the 5\% level. $\Delta$ and the gap are rounded independently; the gap, CI and $p$ are computed on unrounded values, so column differences may differ by up to 0.001.}
  \label{tab:static_vs_dynamic}
  \begin{tabular*}{\textwidth}{@{\extracolsep{\fill}}lccr@{\,}lc@{}}
    \toprule
    \textbf{Method} & $\Delta_{\text{S}}$ & $\Delta_{\text{D}}$ & \multicolumn{2}{c}{\textbf{Gap [95\% CI]}} & \textbf{$p$} \\
    \midrule
    TASH-Index & $\downarrow$0.075 & $\downarrow$0.059 & \textbf{-0.016} & \textbf{[-0.024, -0.009]} & \textbf{$<$0.001} \\
    Random Forest & $\downarrow$0.074 & $\downarrow$0.066 & \textbf{-0.009} & \textbf{[-0.016, -0.002]} & \textbf{0.050} \\
    Repost Count & $\downarrow$0.006 & $\downarrow$0.005 & -0.000 & [-0.009, +0.008] & 0.715 \\
    Coordination Centrality & $\downarrow$0.003 & $\downarrow$0.004 & +0.002 & [-0.011, +0.012] & 0.119 \\
    \bottomrule
  \end{tabular*}
\end{table}

Two findings stand out.
First, for every method, static evaluation reports a reduction at least as large as the dynamic one (gap $\leq 0$ or null): it never \emph{significantly under}-states effectiveness, so it acts as an upper bound.
Second, and crucially, the overestimation is statistically significant \emph{precisely for the two effective rankers}, TASH-Index (gap $-0.016$, $p<0.001$) and Random Forest (gap $-0.009$, $p=0.05$), and indistinguishable from zero for the two ineffective ones (Repost Count, Coordination Centrality; $p>0.1$).
In relative terms, real-time adaptation erases about 21\% of TASH-Index's and 12\% of Random Forest's predicted reduction.
Before the moderation day, the paired gap is statistically zero for all methods (a balance check); it departs from zero only afterwards.

\begin{figure}[tb]
\centering
\includegraphics[width=\textwidth]{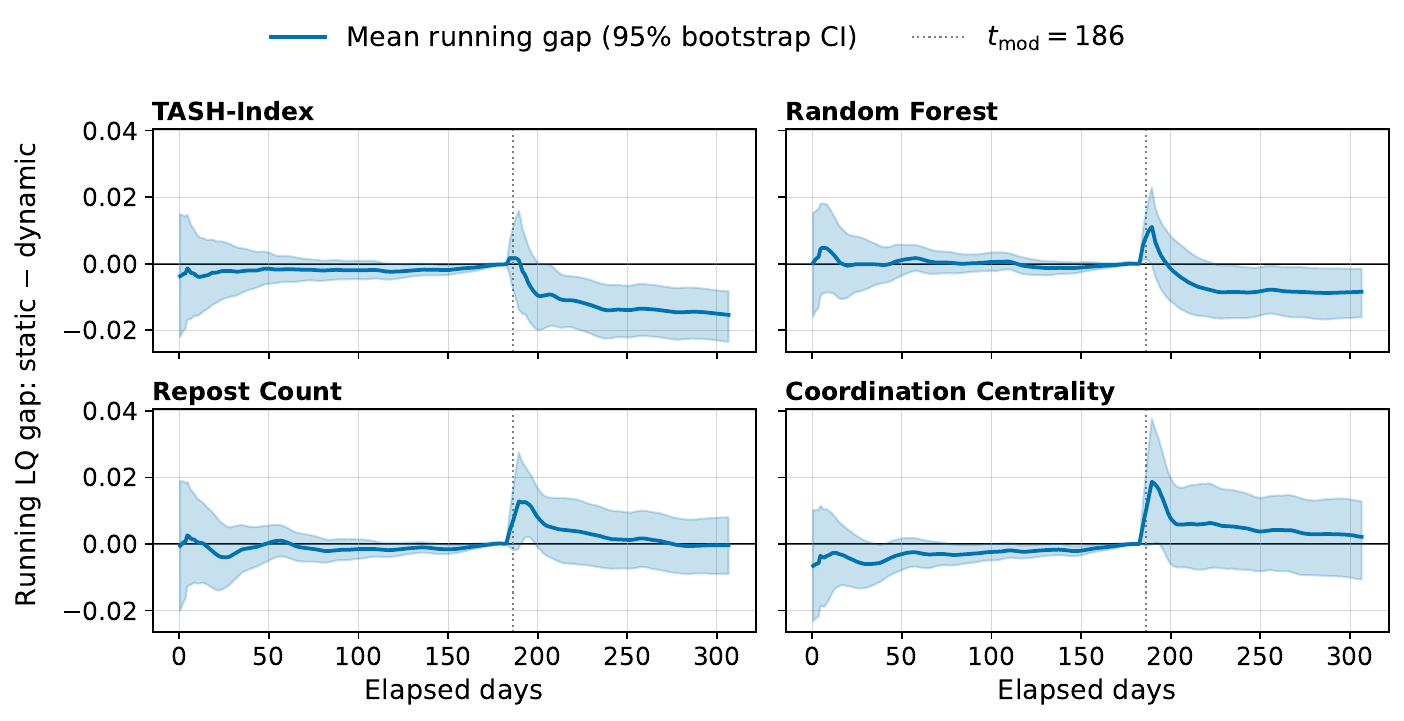}
\caption{%
Running within-network static $-$ dynamic gap in daily LQ fraction per ranking method (mean over the 30 networks, 95\% bootstrap CI; dotted line: $t_{\mathrm{mod}}=186$). Referenced to each network's pre-moderation baseline, the gap is zero before moderation and afterwards converges to the paired gap of Table 2. Negative values mean static evaluation overestimates real-time effectiveness. Only TASH-Index and Random Forest separate from zero; the other two rankers stay within the 95\% CI of zero.}
\label{fig:static_vs_dynamic}
\end{figure}

\paragraph{Interpretation over time.}
Figure~\ref{fig:static_vs_dynamic} plots the \emph{running} static$-$dynamic gap over time. Before the moderation day, it hovers at zero for every method, a balance check confirming the two conditions are equivalent on the same network; after the ban, it separates from zero only for the effective rankers (TASH-Index, Random Forest), settling at a negative value, while staying at zero for the ineffective ones. A negative gap means static evaluation reports a larger LQ reduction than the live simulation delivers: once top spreaders are removed, the remaining users partially compensate by resharing other LQ sources, a feedback invisible to static retrospective removal. Because the difference is taken within each network, the between-network variance cancels, and the band is tight.

\paragraph{Network-level comparison.}
The same pattern holds, more sharply, at the network level: for the effective rankers static evaluation overestimates the reduction in LQ-reshare edges and total weight by 18--25~pp (e.g.\ TASH-Index edges: $\downarrow$36\% static vs.\ $\downarrow$11\% dynamic), whereas for the ineffective rankers the gaps stay within a few percentage points with no qualitative reversal, and the largest weakly connected component barely moves under any regime.

\section{Discussion}
\label{sec:discussion}

\paragraph{Validity and non-circularity of the calibrated SimSoM.}
Principled parameter fitting (GB2, CMA-ES, grid search) yields a simulator that closely reproduces real dynamics. A natural concern is circularity: do the results merely reflect baked-in assumptions? We address this in three ways. First, the quality distribution is fitted to the empirical credibility distribution, reproducing observed proportions rather than arbitrary choices. Second, the \emph{identity} of top spreaders is not predetermined but \emph{emerges} from activity, network position, appeal, and feed dynamics. Third, the targeted-removal validation (RQ2) shows that the same ranking methods identify comparably influential users in real and simulated data, so the emergent spreading structure matches reality.

\paragraph{Static evaluation as an upper bound.}
The overestimation by static analysis (RQ3) has direct policy implications: decision-makers relying on retroactive evaluation may over-credit planned interventions. The gap is significant exactly for the rankers a practitioner would deploy and null for the ineffective ones, which pre-empts the concern that it might matter only for poor detectors. The results suggest a compensatory resharing mechanism: the remaining users fill the void left by banned spreaders, an adaptation invisible to static removal that complements prior work on moderation effectiveness~\cite{schneider2023moderation} and friction interventions~\cite{jahn2025friction}. Static evaluation should therefore be read as an \emph{upper bound} on real-time effectiveness, loosest for the strong detectors that matter most in practice.

\paragraph{Limitations and future work.}
The simulator does not model exogenous events, platform algorithm changes, or user creation/deletion, and its networks (50k--90k nodes) are smaller than major platforms. The grounding is specific to one platform (Twitter), language (Italian), topic (COVID-19 vaccines), and window (2020--2021), so the calibrated parameters and absolute effect sizes should transfer with caution; the \emph{methodology} and the \emph{qualitative} compensatory mechanism, however, rest on generic properties of feed-based networks (heavy-tailed activity, redundant resharing paths) and should recur elsewhere, with magnitude varying by network structure, moderation latency, and cultural context.
Future work should test other interventions (labelling, friction, downranking) and study how compensatory dynamics scale across networks and platforms.

\section{Conclusion}
\label{sec:conclusion}

We presented a calibrated, validated extension of SimSoM and used it as a moderation testbed on VaccinItaly. Beyond reproducing key statistical signatures of real dynamics (RQ1), the simulator reproduces the observed effectiveness of four independent moderation strategies (RQ2), showing that it captures the influence structure behind intervention outcomes rather than merely matching aggregate distributions. This intervention-level validity is what licenses the central finding: when moderation is applied dynamically, static evaluation significantly overestimates the effectiveness of user bans (RQ3). More broadly, once a simulator reproduces observed moderation outcomes, it can probe adaptive effects and counterfactual policies inaccessible to retrospective analysis.

\section*{Acknowledgments}
\label{sec:ack}

This work was partially supported by the Swiss National Science Foundation (grant number CRSII5\_209250).

\begin{credits}
\subsubsection{\discintname}
The authors have no competing interests to declare that are relevant to the content of this article.
\end{credits}

\bibliographystyle{splncs04}
\bibliography{references}

\end{document}